\documentclass[conference]{IEEEtran}
\usepackage{url}
\usepackage{amsmath}
\usepackage[table]{xcolor}
\usepackage{array}
\usepackage{graphicx}
\usepackage{float}
\newfloat{algorithm}{tbp}{loa}
\floatname{algorithm}{Algorithm}

\definecolor{gaintext}{RGB}{38,128,66}
\definecolor{misstext}{RGB}{180,52,42}

\makeatletter
\long\def\@makecaption#1#2{%
\ifx\@captype\@IEEEtablestring%
\footnotesize\bgroup\par\centering\@IEEEtabletopskipstrut{\normalfont\footnotesize #1}\\{\normalfont\footnotesize #2}\par\addvspace{0.5\baselineskip}\egroup%
\@IEEEtablecaptionsepspace
\else
\@IEEEfigurecaptionsepspace
\setbox\@tempboxa\hbox{\normalfont\footnotesize {#1.}\nobreakspace\nobreakspace #2}%
\ifdim \wd\@tempboxa >\hsize%
\setbox\@tempboxa\hbox{\normalfont\footnotesize {#1.}\nobreakspace\nobreakspace}%
\parbox[t]{\hsize}{\normalfont\footnotesize\noindent\unhbox\@tempboxa#2}%
\else%
\hbox to\hsize{\normalfont\footnotesize\hfil\box\@tempboxa\hfil}%
\fi\fi}
\makeatother

\makeatletter
\let\custodyoldcaption\@makecaption
\long\def\@makecaption#1#2{%
\ifx\@captype\@IEEEtablestring
\par\centering{\normalfont\footnotesize #1.\enspace #2\par}\vspace{5pt}%
\else\custodyoldcaption{#1}{#2}\fi}
\makeatother

\usepackage{eso-pic}
\AddToShipoutPictureBG*{\AtPageLowerLeft{\put(36,18){\parbox{540pt}{\fontsize{6}{7}\selectfont
\copyright\ 2026 IEEE. Personal use of this material is permitted. Permission from IEEE must be obtained for all other uses, including reprinting or republishing this material for advertising or promotional purposes, creating new collective works, for resale or redistribution to servers or lists, or reuse of any copyrighted component of this work in other works.}}}}

\begin{document}
\raggedbottom

\title{Same Data, Different Bytes:\\Serialization-Invariant Integrity Checks\\for Agent-Published Datasets}

\author{\IEEEauthorblockN{Jin Gao}
\IEEEauthorblockA{Independent Researcher, USA\\gaojin@alum.mit.edu}
\and
\IEEEauthorblockN{Maura Clark}
\IEEEauthorblockA{UOR Foundation, USA\\maura@uor.foundation}}

\maketitle

\begin{abstract}
Agent networks require reliable ways for requesting agents to verify where data is located, how it can be accessed, and whether the data received matches what the data owner made available.
Checksum mismatches can reflect either altered content or serialization differences that preserve data equivalence under specified rules.
Data identification mechanisms for agents, such as DataFacts~\cite{gao2026datafacts}, describe where data can be accessed and provide checksums for checking received data against published references.
We propose a serialization-invariant identification method that assigns the same identifier to data representations equivalent under explicit rules.
We use DataFacts as a case study, retaining its existing checksum alongside the proposed identifier.
In a controlled agent study, we compare the proposed configuration with ordinary-tool and checksum baselines across equivalent re-encodings, altered content, and unavailable references.
The study includes three model tiers from each of two providers.
For equivalent re-encodings, the proposed configuration increased expected-outcome rates by 80.4 percentage points over the ordinary-tool baseline and by 70.2 percentage points over the checksum baseline.
The proposed configuration also increased evidence-supported rejection of altered content by 3.5 percentage points over the checksum baseline.
We contribute a serialization-invariant identification method with explicit equivalence rules, an integration approach illustrated through DataFacts, and an empirical evaluation against ordinary-tool and checksum baselines.
\end{abstract}


\begin{figure*}[t]
\centering
\includegraphics[width=\textwidth]{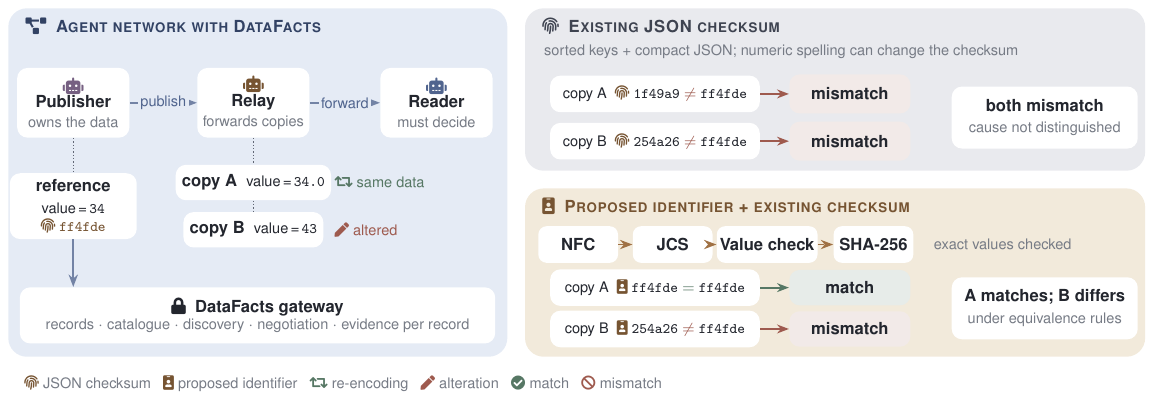}
\caption{Illustrative DataFacts exchange. Re-encoding a numeric value from 34 to 34.0 changes the existing JSON checksum but preserves the proposed identifier; changing the value to 43 changes both. Hashes are shown as six-character prefixes.}
\label{fig:identity}
\end{figure*}

\section{Introduction}

Reliable data exchange in agent networks requires requesting agents to locate relevant data, determine how to access it, and verify that the data received corresponds to what the data owner made available.
DataFacts~\cite{gao2026datafacts} is one example of a data layer that supports these needs: its records describe datasets and their access conditions, and include a checksum computed from a serialized representation of the content.

DataFacts computes its checksum by parsing the JSON content, sorting object member names, and serializing the result without extra whitespace before hashing.
During agent-mediated data exchange, agent-generated outputs, intermediary processing, or serialization tools may change how values are represented---for example, writing the same numeric value as \texttt{1} rather than \texttt{1.0}, or using different Unicode forms for canonically equivalent text---resulting in different checksums.

Because a checksum mismatch can result from either a change in representation or a change in data under the defined equivalence relation, the mismatch alone does not tell a receiving agent whether to accept or reject the data.
Fig.~\ref{fig:identity} illustrates an intermediary agent changing the representation of a numeric value from \texttt{34} to \texttt{34.0}, producing an equivalent copy with a different checksum.
Both the re-encoded copy~A and the altered copy~B produce checksum mismatches.
Applying the same canonicalization rules to the reference dataset and each copy yields matching identifiers for A and the reference, but different identifiers for B in this example.

\textbf{This paper.} We propose a validation workflow for agent networks that accommodates representation differences, checks numeric preservation, and supplies evidence for agent decisions.
Using DataFacts as a case study, we integrate an identifier supporting these checks alongside its existing checksum while retaining its discovery, access-control, and negotiation mechanisms.
We evaluate the method in a small agent network running DataFacts, comparing verification outcomes across equivalent re-encodings, altered content, and unavailable references.

\section{Related Work}
\emph{Agent Data Infrastructure.} DCAT provides a foundation for machine-readable dataset descriptions~\cite{dcat3}, while recent work examines how semantic metadata supports data retrieval by agents~\cite{chen2026metadata}.
DataFacts connects agent discovery to dataset access through metadata describing available data, access conditions, and checksums~\cite{gao2026datafacts}.
Our work complements these discovery and access mechanisms by checking whether received data remain equivalent to a specified reference despite changes in serialization.

\emph{Content Identity.} Canonical representations provide a basis for comparing differently serialized data: the JSON Canonicalization Scheme (JCS) defines deterministic serialization rules for JSON, including object member ordering and numeric representation~\cite{rfc8785}.
JCS preserves Unicode strings without normalization; Unicode Normalization Form C (NFC) provides a complementary rule for mapping canonically equivalent character sequences to a common representation~\cite{rfc8785,uax15}.
Building on these techniques, we study how an explicit equivalence relation can support data verification in agent networks, using DataFacts to examine integration and agents' verification decisions.

\emph{Data Verification.} Digest-based verification is supported by HTTP Digest Fields~\cite{rfc9530}, while RDF Dataset Canonicalization and Trusty URIs enable comparison or verification beyond a particular serialization~\cite{rdfcanon,trustyuris}.
Building on this foundation, we examine whether agents can use explicitly defined data-equivalence checks to distinguish equivalent copies from altered content and recognize when reference evidence is unavailable.

\section{Background: Data Representation and Verification}
\label{sec:background}
JSON separates structured values from their serialized representations: different byte sequences can encode the same value.
For example, representations may differ in object member order, whitespace, numeric notation (e.g., \texttt{1}, \texttt{1.0}, or \texttt{1e0}), or character escaping (e.g., \texttt{"a"} versus \texttt{"\textbackslash u0061"}), without changing the represented values.
Canonically equivalent Unicode strings may additionally use different code-point sequences; treating these as equivalent requires an explicit normalization rule.
Such representation differences can arise when agents generate outputs or intermediaries parse and reserialize data during exchange.

A checksum is computed from a specific byte representation of the data, so changes in that representation can produce different checksums for data that remain equivalent under the chosen rules.
When two checksums differ, the receiving agent cannot determine from that difference alone whether the data have changed or only their representation has changed.
The verification procedure converts received data into a common representation under explicit comparison rules and computes an identifier from that representation, without requiring the sending agent to produce a fixed serialization.
Inputs that cannot be processed under these rules are reported as unsupported or invalid, rather than treated as evidence of a data mismatch.
The rules allow differences in formatting and equivalent ways of writing numbers or text, but treat changes to array order, data types, or values as changes to the data.
Data outside the supported input range may require additional preparation before this check can be used.

\section{A Serialization-Invariant Identification Method}
\label{sec:identifier}
\subsection{Comparison Rules and Identifier Computation}
In agent networks, data exchange involves a data owner providing a dataset and its checksum, intermediary agents or services forwarding the data in potentially different representations, and a requesting agent checking the received copy before use.
The verification method applies the same processing rules to the data owner's reference data and the received copy, then compares the resulting identifiers.
For supported inputs, we compare numbers by exact decimal value and compare strings and object member names after NFC normalization.
Reordering fields in an object preserves equivalence, but reordering array elements or changing a number into a string changes the data.
Inputs must be valid UTF-8 JSON with valid Unicode strings and no duplicate object member names before or after NFC normalization.

Specifically, the procedure validates and parses the JSON input, applies Unicode NFC to strings and object member names~\cite{uax15}, serializes the result using JCS (RFC~8785)~\cite{rfc8785}, and computes its SHA-256 hash as the identifier (Algorithm~\ref{alg:validation}).
We parse each JSON number directly from its original text into an exact decimal value.
For JCS serialization, \texttt{PrepareNumbers} checks numeric bounds and converts exact decimal values into integers or finite binary64 numbers.
After JCS serialization, we parse the output again and verify that every number has exactly the same decimal value as in the input.
\texttt{EqualExact} compares objects by member name, arrays by position, and scalar values by type and value, using exact decimal equality for numbers.
We compute the SHA-256 identifier from the serialized bytes after this check passes.
For example, representing a field's numeric value as \texttt{0.1}, \texttt{0.10}, or \texttt{1e-1} yields the same identifier when all other data are unchanged.
\texttt{Classify} reports precision, range, and resource-limit failures as unsupported input, and other input failures as invalid input.

\begin{algorithm}[t]
\refstepcounter{algorithm}
\centering
\begin{minipage}{\columnwidth}
\fontsize{8}{9.5}\selectfont
\noindent\textbf{Algorithm \thealgorithm. Serialization-invariant JSON validation}\par
\vspace{3pt}\hrule\vspace{4pt}
\noindent\textbf{Input:} Original JSON bytes $B$; owner's reference record $R$.\par
\noindent\textbf{Output:} Status, comparison evidence, or failure reason.\par
\vspace{3pt}
\setlength{\tabcolsep}{2pt}
\renewcommand{\arraystretch}{1.05}
\begin{tabular}{@{}r@{\hspace{5pt}}p{\dimexpr\columnwidth-16pt\relax}@{}}
1 & $q \gets \operatorname{CheckReference}(R)$ \\
2 & \textbf{if} $q$ failed: \textbf{return} its status and reason \\
3 & \textbf{try} \\
4 & \quad $x \gets \operatorname{ParseExact}(B)$ \\
5 & \quad $y \gets \operatorname{NFCRecursive}(x)$ \\
6 & \quad $b \gets \operatorname{JCS}(\operatorname{PrepareNumbers}(y))$ \\
7 & \quad $z \gets \operatorname{ParseExact}(b)$ \\
8 & \quad \textbf{if not} $\operatorname{EqualExact}(y,z)$: \\
9 & \qquad \textbf{raise} \texttt{NUMBER\_PRECISION\_LOSS} \\
10 & \quad $h \gets \operatorname{SHA256}(b)$ \\
11 & \textbf{catch} processing failure $e$: \\
12 & \quad \textbf{return} $\operatorname{Classify}(e)$ and its reason \\
13 & $h_{\mathrm{ref}} \gets R.\mathrm{identifier}$ \\
14 & $s \gets \texttt{match}$ if $h=h_{\mathrm{ref}}$; else \texttt{mismatch} \\
15 & \textbf{return} $(s,h_{\mathrm{ref}},h)$ with reference and profile \\
\end{tabular}
\vspace{4pt}\hrule
\end{minipage}

\label{alg:validation}
\end{algorithm}

\subsection{DataFacts Integration and Verification}
In the DataFacts integration, the identifier is computed solely from the dataset content; descriptive metadata, access settings, and the identifier itself are excluded from the calculation.
In the DataFacts implementation, the gateway computes the proposed identifier from the stored dataset content and publishes it alongside the existing checksum in the record's evidence fields.
The published record also identifies the processing profile, allowing the verifier to check that both identifiers were computed using the same procedure.
If this profile is missing or incompatible, the verifier reports that the comparison cannot be completed.
The receiving agent computes an identifier from the received data and checks whether the computed identifier matches the identifier published by the data owner.
The identifier accompanying a forwarded copy is not sufficient on its own, because the copy and its identifier could both have been replaced.
The comparison assumes that the identifier obtained from the data owner's record is authentic.
If the data owner's identifier is unavailable, the receiving agent reports that the data cannot be verified.

The existing DataFacts checksum already ignores object member order and extra whitespace, but can change when equivalent numbers or text use different representations.
Our proposed method applies explicit rules to make the identifier invariant to these supported representation differences, while retaining the same recompute-and-compare verification procedure.
DataFacts continues to enforce its existing access permissions; the proposed check determines whether the received content matches the owner's data under the specified comparison rules.

\section{Evaluation}
\subsection{Experimental Setup}
We compare the three verification configurations on 36 JSON materials covering both integer and fractional values.
For each material, we test three exchange scenarios: equivalent re-encoding, altered content, and an unavailable reference from the data owner (Table~\ref{tab:exchange-cases}).
Each task asks the agent to verify the entire forwarded dataset against the data owner's record before using it to answer a data-dependent question.
In the equivalent and altered-content scenarios, agents can access the data owner's reference metadata but cannot retrieve the owner's original content.
In the unavailable-reference scenario, the candidate data are unchanged from the equivalent re-encoding scenario, but access to the data owner's reference record is denied.
For each material, we use the same task prompt across scenarios and configurations, without revealing the scenario or expected decision.

The ordinary-tool baseline provides existing DataFacts metadata and general-purpose tools for inspecting the received data, without a dedicated integrity verifier.
The checksum baseline provides the same metadata and general-purpose tools, together with a verifier that compares the received data's existing JSON checksum with the checksum in the data owner's reference record.
The proposed configuration retains the existing checksum and compares the received data's serialization-invariant identifier with the identifier in the data owner's reference record.

We test gpt-5.6-luna, gpt-5.6-terra, gpt-5.6-sol, claude-haiku-5-5, claude-sonnet-5-5, and claude-opus-5-5, repeating each combination of material, scenario, model, and configuration three times.
Trials run in isolated Docker agent deployments with A2A communication and authenticated DataFacts gateway access; JCS serialization uses rfc8785 version 0.1.4.
We require each response to include a decision (accept, reject, or unable to verify), a brief explanation, and an answer only when the data are accepted.
The expected outcome is acceptance of equivalent data, evidence-supported rejection of altered content, or evidence-supported abstention when the data owner's reference is unavailable.
For equivalent data, acceptance counts toward the expected-outcome rate only when the agent also returns the expected answer.
We count a rejection as evidence-supported only when the recorded tool observations show a mismatch with the data owner's reference.
An unable-to-verify decision requires a recorded failed attempt to access the data owner's reference record, with no observed match or mismatch.
Format-invalid responses received at most one replacement, irrespective of outcome correctness; unreplaced invalid responses counted as unsuccessful.

\begin{table}[t]
\centering
\caption{Exchange scenarios and test cases.}
\label{tab:exchange-cases}
\begingroup
\fontsize{8}{9}\selectfont
\setlength{\tabcolsep}{3pt}
\renewcommand{\arraystretch}{1.0}
\begin{tabular}{@{}>{\raggedright\arraybackslash}p{\dimexpr .22\columnwidth-4pt\relax}>{\raggedright\arraybackslash}p{\dimexpr .54\columnwidth-4pt\relax}>{\raggedright\arraybackslash}p{\dimexpr .24\columnwidth-4pt\relax}@{}}
\hline
\textbf{Case} & \textbf{Example} & \textbf{Expected} \\
\hline
\rowcolor{black!7}\multicolumn{3}{@{}l@{}}{\textbf{Equivalent re-encoding}} \\
Formatting & Change whitespace or key order & Accept \\
Integers & \texttt{64} vs. \texttt{64.0} & Accept \\
Trailing zeros & \texttt{21.5} vs. \texttt{21.500} & Accept \\
Exponents & \texttt{-12.75} vs. \texttt{-1275e-2};\newline \texttt{0.0000001} vs. \texttt{1e-7} & Accept \\
Fractions & \texttt{0.1} vs. \texttt{1e-1} & Accept \\
Unicode & Unicode composition in text/keys & Accept \\
JSON escapes & Literal vs. escaped characters & Accept \\
Objects/arrays & Object-key order; fixed array order & Accept \\
Combined & Order, numbers, and Unicode & Accept \\
\rowcolor{black!7}\multicolumn{3}{@{}l@{}}{\textbf{Altered content}} \\
Values & \texttt{0.1} vs. \texttt{0.1001}; remove an accent & Reject \\
Types & \texttt{7} vs. \texttt{"7"} & Reject \\
Structure & Field location; array order or size & Reject \\
Other fields & Change a non-queried field & Reject \\
\rowcolor{black!7}\multicolumn{3}{@{}l@{}}{\textbf{Unavailable reference}} \\
Reference & Owner reference inaccessible & \mbox{Unable to verify} \\
\hline
\end{tabular}
\endgroup
\end{table}

\begin{table}[t]
\centering
\caption{Expected-outcome rates by scenario.}
\label{tab:scenario-baseline-comparison}
\begingroup
\definecolor{outcomegreen}{RGB}{20,92,48}
\definecolor{outcomered}{RGB}{145,30,38}
\fontsize{8}{9}\selectfont
\setlength{\tabcolsep}{3pt}
\renewcommand{\arraystretch}{1.16}
\begin{tabular*}{\columnwidth}{@{\extracolsep{\fill}}lrrr@{}}
\hline
\textbf{Comparison baseline} & \textbf{Baseline rate (\%)} & \textbf{Proposed rate (\%)} & \textbf{Gain (pp)} \\
\hline
\rowcolor{black!7}\multicolumn{4}{@{}l@{}}{\textbf{Equivalent re-encoding (accept)}} \\
Ordinary tools & 18.4 & 98.8 & \textcolor{outcomegreen}{+80.4} \\
Checksum verifier & 28.5 & 98.8 & \textcolor{outcomegreen}{+70.2} \\
\rowcolor{black!7}\multicolumn{4}{@{}l@{}}{\textbf{Altered content (evidence-supported rejection)}} \\
Ordinary tools & 65.9 & 97.4 & \textcolor{outcomegreen}{+31.5} \\
Checksum verifier & 93.8 & 97.4 & \textcolor{outcomegreen}{+3.5} \\
\rowcolor{black!7}\multicolumn{4}{@{}l@{}}{\textbf{Unavailable reference (evidence-supported abstention)}} \\
Ordinary tools & 99.2 & 99.5 & \textcolor{outcomegreen}{+0.3} \\
Checksum verifier & 99.5 & 99.5 & 0.0 \\
\hline
\end{tabular*}
\endgroup
\end{table}

\begin{table}[t]
\centering
\caption{Expected-outcome rates by model and scenario (\%).}
\label{tab:model-scenario-comparison}
\begingroup
\fontsize{7.5}{8.5}\selectfont
\setlength{\tabcolsep}{2pt}
\renewcommand{\arraystretch}{1.08}
\begin{tabular*}{\columnwidth}{@{\extracolsep{\fill}}lrrrrrrr@{}}
\hline
& \multicolumn{3}{c}{\textbf{GPT}} & \multicolumn{3}{c}{\textbf{Claude}} & \\
\cline{2-4}\cline{5-7}
\textbf{Configuration} & \textbf{Luna} & \textbf{Terra} & \textbf{Sol} & \textbf{Haiku} & \textbf{Sonnet} & \textbf{Opus} & \textbf{All} \\
\hline
\rowcolor{black!7}\multicolumn{8}{@{}l@{}}{\textbf{Equivalent re-encoding}} \\
Ordinary tools & 42.6 & 28.7 & 33.3 & 4.6 & 0.9 & 0.0 & 18.4 \\
Checksum & 44.4 & 43.5 & 42.6 & 28.7 & 12.0 & 0.0 & 28.5 \\
Proposed & 99.1 & 100.0 & 100.0 & 99.1 & 100.0 & 94.4 & 98.8 \\
\rowcolor{black!7}\multicolumn{8}{@{}l@{}}{\textbf{Altered content}} \\
Ordinary tools & 25.0 & 63.0 & 81.5 & 28.7 & 98.1 & 99.1 & 65.9 \\
Checksum & 79.6 & 97.2 & 96.3 & 91.7 & 100.0 & 98.1 & 93.8 \\
Proposed & 95.4 & 98.1 & 98.1 & 92.6 & 100.0 & 100.0 & 97.4 \\
\rowcolor{black!7}\multicolumn{8}{@{}l@{}}{\textbf{Unavailable reference}} \\
Ordinary tools & 100.0 & 99.1 & 99.1 & 97.2 & 100.0 & 100.0 & 99.2 \\
Checksum & 100.0 & 98.1 & 100.0 & 99.1 & 100.0 & 100.0 & 99.5 \\
Proposed & 100.0 & 97.2 & 100.0 & 100.0 & 100.0 & 100.0 & 99.5 \\
\hline
\end{tabular*}
\endgroup
\end{table}

Offline checks confirmed that decimal values such as 0.1 remain JSON numbers, while inputs that would lose precision during serialization are reported as unsupported.

\subsection{Results}
For equivalent re-encodings, the proposed configuration achieved the expected acceptance outcome in 98.8\% of trials, exceeding the ordinary-tool baseline by 80.4 percentage points and the checksum baseline by 70.2 percentage points (Table~\ref{tab:scenario-baseline-comparison}).
For altered content, the proposed configuration achieved evidence-supported rejection in 97.4\% of trials, 3.5 percentage points above the checksum baseline.
When the data owner's reference was unavailable, both the proposed configuration and the checksum baseline achieved evidence-supported abstention in 99.5\% of trials.
Model-level gains in acceptance of equivalent data ranged from 56.5 to 99.1 percentage points over the ordinary-tool baseline and from 54.6 to 94.4 percentage points over the checksum baseline (Table~\ref{tab:model-scenario-comparison}).
These results show that the proposed method improves data validation across models from both providers, particularly when equivalent data are expressed in different forms.

The largest gains occur when agents receive equivalent data in a different representation, whereas the checksum baseline already achieves high expected-outcome rates for altered content and unavailable references.
Adding the checksum verifier to ordinary tools increased acceptance of equivalent data by 10.2 percentage points.
Both verifier configurations provide an explicit verification action; their comparison evaluates differences in the comparison procedure and the evidence returned to agents.
Non-expected outcomes involved agents rejecting or withholding judgment despite matching evidence, withholding judgment despite mismatch evidence, or failing to provide the expected answer or required candidate content.

\section{Discussion and Conclusion}
\label{sec:limits}
\emph{Discussion.} The method supports validation of structured JSON in agent networks by allowing explicitly defined representation differences while checking that the data remain equivalent to a reference.
This includes JSON generated directly by agents, where formatting or object field order may vary without changing the represented values.
The method gives agents flexibility in how they represent JSON while retaining strict checks on data equivalence.

Full-content validation can detect changes beyond the fields needed to answer the immediate task.
Applications should use equivalence rules that preserve the distinctions relevant to their tasks, such as the number of decimal places when it conveys measurement precision.
The same validation approach can support other agent workflows that provide a trusted reference identifier for the JSON content being checked.

\emph{Limitations.} The evaluation uses selected JSON transformations in a small, controlled agent network; naturally occurring agent outputs, broader deployments, and other data formats remain to be evaluated.
The implementation supports numeric inputs only when serialization preserves their exact decimal values; inputs that would lose precision are reported as unsupported.

In conclusion, we propose a serialization-invariant validation method for agent networks, using DataFacts as an integration and evaluation case.
The controlled experiments show large gains over both baselines in accepting equivalent data, along with improved evidence-supported rejection of altered content.
This gives agent networks a flexible way to exchange structured data while retaining rigorous validation against a reference.

\section*{AI Use Disclosure}
AI assistants were used for code development, data analysis, and manuscript drafting.
The authors are responsible for all reported results and citations.

\end{document}